\documentclass[aps,nofootinbib,prd,eqsecnum,twocolumn,showpacs,showkeys,preprintnumbers]{revtex4-1}
\usepackage{graphicx}
\usepackage{lineno}
\usepackage{graphicx}
\usepackage{amsmath}
\usepackage{amsfonts}
\usepackage{amssymb}
\usepackage{color}
\usepackage{bm}
\usepackage{float}
\usepackage{mathrsfs}
\usepackage{epstopdf}
\usepackage{url}
\usepackage{placeins}
\usepackage{footnote}
\usepackage{textcomp}
\usepackage[normalem]{ulem}
\usepackage{xcolor}
\usepackage[unicode=true, pdfusetitle,
 bookmarks=true,bookmarksnumbered=false,bookmarksopen=false,
 breaklinks=false,pdfborder={0 0 1},backref=false,colorlinks=false]{hyperref}
\usepackage{multirow}
\usepackage{pifont}
\usepackage{times}
\usepackage[english]{babel}

\usepackage{adjustbox}
\usepackage{amsmath}
\usepackage{booktabs}
\usepackage{caption}

\usepackage{float}
\usepackage{enumerate}
\usepackage{lineno}
\usepackage{hyperref}

\usepackage{tabularx}
\makeatletter

\newcommand{\stkout}[1]{\ifmmode\text{\sout{\ensuremath{#1}}}\else\sout{#1}\fi}

\newcolumntype{L}[1]{>{\hsize=#1\hsize\raggedright\arraybackslash}X}%
\newcolumntype{R}[1]{>{\hsize=#1\hsize\raggedleft\arraybackslash}X}%
\newcolumntype{C}[1]{>{\hsize=#1\hsize\centering\arraybackslash}X}%

\newcommand*\patchAmsMathEnvironmentForLineno[1]{%
 \expandafter\let\csname old#1\expandafter\endcsname\csname #1\endcsname
 \expandafter\let\csname oldend#1\expandafter\endcsname\csname end#1\endcsname
 \renewenvironment{#1}%
   {\linenomath\csname old#1\endcsname}%
   {\csname oldend#1\endcsname\endlinenomath}}%
\newcommand*\patchBothAmsMathEnvironmentsForLineno[1]{%
 \patchAmsMathEnvironmentForLineno{#1}%
 \patchAmsMathEnvironmentForLineno{#1*}}%
\AtBeginDocument{%
\patchBothAmsMathEnvironmentsForLineno{align}%
\patchBothAmsMathEnvironmentsForLineno{flalign}%
\patchBothAmsMathEnvironmentsForLineno{alignat}%
\patchBothAmsMathEnvironmentsForLineno{gather}%
\patchBothAmsMathEnvironmentsForLineno{multline}%
}

\begin{document}
\title{RSD constraints on power-law $f(Q)$ gravity using Barboza-Alcaniz and Jassal-Bagla-Padmanabhan parametrizations}
\author{Chaymae Karam $^{1}$}
\email{chaymae$_$karam2@um5.ac.ma}
\author{Dalale Mhamdi$^{2,3}$}
\email{dalale.mhamdi@ump.ac.ma}
\author{Taoufik Ouali$^{2,3}$}
\email{t.ouali@ump.ac.ma}
\author{Rachid Ahl Laamara$^{1}$}
\email{r.ahllaamara@um5r.ac.ma}
\author{Mohamed Bennai$^{1,4}$}
\email{mohamed.bennai@univh2c.ma}
\date{\today }

\affiliation{$^{1}$Laboratory of High Energy Physics, Modeling and Simulations, Faculty of Science, University Mohammed V in Rabat, Rabat, Morocco.\\$^{2}$Laboratory of Physics of Matter and Radiation, University of Mohammed first, BP 717, Oujda, Morocco.\\$^{3}$Astrophysical and Cosmological Center, Faculty of Sciences, University of Mohammed first, BP 717, Oujda, Morocco.\\$^{4}$Quantum Physics and Spintronic Team, LPMC, Faculty of Sciences Ben M’sick, Hassan II University of Casablanca, Morocco.
\\}


\begin{abstract}
We investigate the late-time accelerated expansion of the Universe in power-law $f(Q)$ gravity, $f(Q) = Q + 6\gamma H_0^2 \left(\frac{Q}{Q_0}\right)^n$, combined with two dynamical dark energy parametrizations, Barboza-Alcaniz (BA) and Jassal-Bagla-Padmanabhan (JBP), which allow a smooth evolution of the equation of state beyond the standard constant $\omega$ assumption. We derive analytical expressions for the Hubble parameter and the dark energy density, and constrain the model using Pantheon$^+$ Type Ia Supernovae, DESI DR2 Baryon Acoustic Oscillations, and Cosmic Chronometers. Going beyond purely geometrical probes, we incorporate Redshift Space Distortion (RSD) measurements to test the growth of cosmic structures and distinguish modified gravity from General Relativity through the effective Newton's constant. Both parametrizations reproduce the observed late-time acceleration and remain consistent with growth data, with only mild deviations from $\Lambda$CDM over low and intermediate redshifts. Notably, both configurations yield $G_{\rm eff} < G$, indicating suppressed growth of matter perturbations. These results establish power-law $f(Q)$ gravity with BA and JBP parametrizations as a viable and versatile alternative to $\Lambda$CDM across both background and perturbative regimes.
\end{abstract}

\maketitle
\section{Introduction}

Over the past decades, a wealth of observational evidence, including Type Ia supernovae~\cite{ref1,ref2}, large-scale structure surveys~\cite{ref3,ref4}, baryon acoustic oscillations~\cite{ref5,ref6}, and measurements of the cosmic microwave background~\cite{ref7,ref8}, has firmly established that the Universe is currently undergoing an accelerated expansion. Within the framework of General Relativity, this late-time acceleration is typically attributed to an additional component with negative pressure, commonly referred to as dark energy~\cite{MAR}.

The standard cosmological model, $\Lambda$CDM, provides a remarkably successful description of these observations by combining a cosmological constant, $\Lambda$, with cold dark matter (CDM). Despite its observational success, this model faces several conceptual difficulties. In particular, the cosmological constant is associated with the well-known fine-tuning and coincidence problems~\cite{ref3,ref4,ref5,ref7,ref8,ref9,ref10,ref11}. In addition, the persistent disagreement between early- and late-time measurements of the Hubble constant, known as the Hubble tension, has emerged as a major challenge, pointing to possible limitations of the standard cosmological scenario~\cite{ref6}.

These issues have motivated the exploration of alternative approaches. One possibility is to consider dynamical forms of dark energy~\cite{NA}, where the equation of state evolves with cosmic time, as in quintessence~\cite{ref12,ref13}, phantom~\cite{ref14,ref15,ref16,ref17,ref18,ref19}, and k-essence models~\cite{ref20, ref21}.Another avenue is to revisit the underlying theory of gravity itself. In General Relativity, gravity is interpreted geometrically as the curvature of space-time, governed by the Ricci scalar, $R$, within a Riemannian framework where both torsion and non-metricity vanish. While this description has been extensively tested and confirmed at local scales, it may not provide a complete account of gravitational phenomena at cosmological distances, especially in light of the observed acceleration and the dynamics of large-scale structures~\cite{BA}.

In this context, modified gravity theories offer a compelling alternative by extending the geometric foundations of General Relativity~\cite{ref23,ref24,ref25,ref26,ref27,ref28,ref29,ref30,ref31,ref32,ref33}. Among these, $f(Q)$ gravity has recently attracted considerable attention~\cite{ref34}. This framework is based on the non-metricity scalar, $Q$, arising in symmetric teleparallel geometry~\cite{FQ}, where gravity is encoded in the non-metricity of spacetime rather than in its curvature or torsion~\cite{PA}. Such an approach provides a geometrically distinct formulation that can naturally accommodate deviations from the standard $\Lambda$CDM model~\cite{de}. In particular, simple power-law forms of $f(Q)$ gravity have been shown to capture a wide range of cosmological behaviors, including late-time acceleration~\cite{ref35,ref36}. Unfortunately, many modify gravities may coincide with each others at the background level in the  homogeneous and isotropic spacetime. Indeed, for instance,  non-metricity and torsion gravities are two equivalent theories at the background level. Since non-metricity and torsion arise from distinct parts of the affine connection, the two theories are not required to predict the same growth of matter perturbations, even on a symmetric background such as Friedmann-Lema\^itre-Robertson-Walker (FLRW)~\cite{PA1}. Distinguishing between these theories therefore requires perturbative probes, such as the RSD measurements described in Sec.~\ref{rsd}.

In parallel, model-independent parametrizations of the dark energy equation of state provide a useful tool to probe deviations from a cosmological constant. Among them, the Barboza-Alcaniz (BA) and Jassal-Bagla-Padmanabhan (JBP) parametrizations are widely used due to their well-behaved over a broad redshift range~\cite{ref37,ref38}. These models allow for a smooth evolution of $\omega(z)$, avoiding divergences and offering a flexible framework to investigate the dynamics of dark energy.

Motivated by these considerations, we explore in this work the cosmological implications of a power law $f(Q)$ gravity model combined with BA and JBP dark energy parametrizations. Specifically, we consider the functional form $f(Q) = Q + 6 \gamma H_0^2 \left(\frac{Q}{Q_0}\right)^n$~\cite{ref39}, where $\gamma$ and $n$ are free parameters, and $Q_0 = 6H_0^2$. This formulation allows for a smooth departure from General Relativity while preserving it as a limiting case.

We derive analytical expressions for the Hubble parameter and the corresponding effective dark energy density for both parametrizations, and investigate their impact on cosmic evolution. Our analysis is constrained using recent observational datasets, including Pantheon$^{+}$ Type Ia Supernovae, DESI DR2 Baryon Acoustic Oscillations, and Cosmic Chronometers, which provide robust probes of the background expansion history across a wide redshift range~\cite{MI}.

To complement these geometrical constraints, we incorporate Redshift Space Distortion (RSD) measurements, which probe the growth rate of large-scale structures~\cite{refF1}. This allows us to extend the analysis to the perturbative level and assess the consistency of the model with structure formation. In modified gravity scenarios, the growth of matter perturbations is typically governed by an effective gravitational coupling, $G_{\mathrm{eff}}$, which may deviate from Newton’s constant $G$. Therefore, RSD data provide a direct way to test such deviations and to distinguish between general relativity and its extensions.

In this work, $G_{\mathrm{eff}}$ serves as a key quantity linking the background dynamics of the $f(Q)$ framework with the evolution of matter perturbations, ensuring a consistent description of both expansion history and structure growth within a unified observational framework.

Our aim is to determine whether this framework can account for the observed late-time acceleration while remaining compatible with both expansion and growth observations, and to identify possible deviations from the standard $\Lambda$CDM scenario. We further compare our model with $\Lambda$CDM using the Akaike and Bayesian information criteria~\cite{ref40,ref41}. To better understand its dynamical features, we analyze several cosmological diagnostics, including the deceleration, jerk, and snap parameters, as well as the $Om(z)$ diagnostic. Our results indicate that these parametrizations within the power law f(Q) framework offer a flexible and comprehensive alternative to conventional dark energy models, capable of describing both the late-time acceleration of the Universe and the evolution of cosmic structures.

The paper is organized as follows. In Sec.~\ref{sec:1}, we outline the theoretical framework of $f(Q)$ gravity. Sec.~\ref{sec:2} is devoted to the dynamical analysis of the power law $f(Q)$ model with BA and JBP parametrizations. In Sec.~\ref{sec:3}, we describe the observational datasets and the statistical methodology. The results and their discussion are given in Sec.~\ref{sec:4}. In Sec.~\ref{sec:5}, we analyze the evolution of key cosmological quantities, including the deceleration, jerk, and snap parameters, as well as the $Om(z)$ diagnostic. Finally, the main conclusions are summarized in Sec.~\ref{sec:6}.

\section{Basic formalism of $f(Q)$ gravity} \label{sec:1}

The $f(Q)$ gravity theory constitutes a geometric extension of the symmetric teleparallel equivalent of general relativity, formulated on a spacetime where both curvature and torsion vanish while gravity is driven by non-metricity \cite{B1,B2}. In this setting, the gravitational interaction is described through the non-metricity scalar, $Q$, which characterizes the variation of the metric under parallel transport. Instead of the Einstein-Hilbert Lagrangian built from the Ricci scalar, $R$, the gravitational sector is generalized by promoting $R$ to an arbitrary function $f(Q)$. This modification provides a flexible framework in which deviations from standard general relativity can be systematically explored. From a cosmological perspective, the $f(Q)$ scenario has attracted considerable attention because it naturally allows for effective dark energy behavior and can reproduce the observed late-time acceleration without introducing additional exotic components. The corresponding gravitational action is expressed as

\begin{equation}
 S=\int\left[\frac{1}{2} f(Q)+\mathcal{L}_m\right] d^4 x \sqrt{-g}                \label{7}
 \end{equation}
 
Here, $\mathcal{L}_m$ represents the matter Lagrangian density, and $g$ denotes the determinant of the metric tensor $g_{\mu \nu}$. We adopt units in which $8 \pi G = 1$. The non-metricity tensor $Q_{\gamma\mu\nu}$ is defined as

\begin{equation}
 \begin{gathered}
Q_{\gamma \mu \nu}=\nabla_\gamma g_{\mu \nu}, \\
Q_\gamma=Q_\gamma{ }^\mu{ }_\mu, \quad \tilde{Q}_\gamma=Q^\mu{ }_{\gamma \mu},
\end{gathered}   
\end{equation} 
here, $\nabla$ denotes the covariant derivative, while the superpotential tensor also referred to as the non-metricity conjugate is defined as
\begin{equation}
4 P^\gamma{ }_{\mu \nu}=-Q^\gamma{ }_{\mu \nu}+2 Q_{(\mu}{ }^\gamma{ }_{\nu)}+Q^\gamma g_{\mu \nu}-\widetilde{Q}^\gamma g_{\mu \nu}-\delta_{(\mu}^\gamma Q_{\nu)},
\end{equation}  
where the trace of the non-metricity tensor can be written as
\begin{equation}
Q=-Q_{\gamma \mu \nu} P^{\gamma \mu \nu}.
\end{equation}\\

By varying the action in Eq.~(\ref{7}) with respect to the metric, one obtains the field equations of $f(Q)$ gravity, written as \cite{B4}
\begin{widetext}
\begin{equation}
\begin{aligned}
    \frac{-2}{\sqrt{-g}} \nabla_\alpha\left(\sqrt{-g} f_Q P_{\mu \nu}^\alpha\right)-\frac{1}{2} g_{\mu \nu} f-f_Q\left(P_{\mu \alpha \beta} Q_\nu^{\alpha \beta}-2 Q_{\alpha \beta \mu} P_\nu^{\alpha \beta}\right)=T_{\mu \nu}
\end{aligned}
\end{equation}
\end{widetext} 

Where the energy-momentum tensor is given by

\begin{equation}
T_{\mu \nu}=-\frac{2}{\sqrt{-g}} \frac{\delta\left(\sqrt{-g} \mathcal{L}_m\right)}{\delta g^{\mu \nu}}. 
\end{equation} 

The variation of the action with respect to the affine connection leads to \cite{B5}
\begin{equation}
\nabla^\mu \nabla^\nu\left(\sqrt{-g} f_Q P^\gamma{ }_{\mu \nu}\right)=0.    
\end{equation}

In $f(Q)$ gravity, the field equations ensure the conservation of the energy-momentum tensor and recover Einstein’s equations in the limit $f(Q)=Q$. For cosmological applications, we consider a spatially flat, homogeneous, and isotropic Universe described by the Friedmann-Lemaître-Robertson-Walker (FLRW) metric

\begin{equation}
ds^2 = -dt^2 + a^2(t)\left(dx^2 + dy^2 + dz^2\right),
\end{equation}
where $t$ is the cosmic time and $a(t)$ the scale factor. Accordingly, the Hubble parameter is defined as
\begin{equation}
H(t) \equiv \frac{\dot{a}}{a}.
\end{equation}

The cosmological redshift is connected to the scale factor via $1+z = \frac{a_0}{a}$. In the context of symmetric teleparallel gravity, the non-metricity scalar corresponding to the FLRW background simplifies to

\begin{equation}
Q = 6H^2.              \label{2,10}
\end{equation}

This relation allows the gravitational sector to be fully formulated in terms of the Hubble parameter and its derivatives, considerably simplifying the cosmological equations. We further assume that the cosmic matter content can be described as a perfect fluid, whose energy-momentum tensor is given by
\begin{equation}
T_{\mu\nu} = (\rho + p)u_\mu u_\nu + pg_{\mu\nu},
\end{equation}
where $\rho$ and $p$ denote the total energy density and pressure, respectively, while $u^\mu$ is the four-velocity of the fluid, normalized such that $u^\mu u_\mu = -1$. 

To incorporate deviations from general relativity, the gravitational Lagrangian is written as
\begin{equation}
f(Q) = Q + F(Q),
\end{equation}
where the function $F(Q)$ accounts for the modifications to the gravitational sector. Substituting the FLRW metric into the field equations of $f(Q)$ gravity yields the generalized Friedmann equations
\begin{equation}
3H^2 = \rho + \frac{F}{2} - QF_Q,
\end{equation}
\begin{equation}
\left(2QF_{QQ} + F_Q + 1\right)\dot{H}
+ \frac{1}{4}\left(Q + 2QF_Q - F\right) = -2p,
\end{equation}
where $F_Q = dF/dQ$ and $F_{QQ} = d^2F/dQ^2$.

The total energy density is decomposed as $\rho = \rho_m + \rho_r$, corresponding to matter and radiation components, respectively. The total pressure reads $p = p_m + p_r$, with the matter pressure being negligible. Each component independently satisfies the standard conservation equation
\begin{equation}
\dot{\rho} + 3H(1+\omega)\rho = 0,
\end{equation}
where $\omega$ represents the equation of state parameter of the corresponding fluid.

\section{Dynamics of Power Law $f(Q)$ Gravity with BA and JBP Parametrizations}
\label{sec:2}

In this section, we explore the cosmological dynamics of the power law $f(Q)$ gravity model by considering two dark energy equation of state (EoS) parametrizations, namely the Barboza-Alcaniz (BA) and the Jassal-Bagla-Padmanabhan (JBP) forms. Unlike constant descriptions, these parametrizations allow the dark energy sector to evolve with redshift and provide a flexible framework for describing late time acceleration.

The BA parametrization is defined as~\cite{ref37}
\begin{equation}
\omega_{de}^{\rm BA}(z)=\omega_0+\omega_1
\frac{z(1+z)}{1+z^2},
\label{eq5BA}
\end{equation}
while the JBP parametrization reads~\cite{ref38}
\begin{equation}
\omega_{de}^{\rm JBP}(z)=\omega_0+\omega_1
\frac{z}{(1+z)^2}.
\label{eq5JBP}
\end{equation}

For a spatially flat FLRW Universe, the modified Friedmann equations take the form
\begin{equation}
3H^2 = \rho_r + \rho_m + \rho_{de},
\label{eq3}
\end{equation}
\begin{equation}
2\dot H + 3H^2 = -\frac13\rho_r - p_{de}.
\end{equation}

Here $\rho_r$ and $\rho_m$ denote the radiation and matter energy densities, whereas $\rho_{\rm de}$ and $p_{\rm de}$ represent the effective dark energy density and pressure emerging from the geometric sector of $f(Q)$ gravity
\begin{align}
\rho_{de} &= \frac{F}{2} - Q F_Q, \label{eq23} \\
p_{de} &= 2\dot H (2QF_{QQ}+F_Q) - \rho_{de}. \label{eq24}
\end{align}

Assuming no interaction between matter, radiation and dark energy, the conservation equations are

\begin{align}
\dot\rho_r + 4H\rho_r &= 0, \label{eq1} \\  
\dot\rho_m + 3H\rho_m &= 0, \label{eq2} \\
\dot\rho_{de} + 3H(1+\omega_{de})\rho_{de} &= 0. \label{eq28}
\end{align}

The effective dark energy equation of state parameter can be written as

\begin{equation}
\omega_{de} = -1 +
\frac{4 \dot H ( 2QF_{QQ} + F_Q )}{F - 2QF_Q}.
\label{eq29}
\end{equation}

From Eqs.~(\ref{eq1}) and (\ref{eq2}), one obtains the standard scaling laws $\rho_m \propto (1+z)^3$ and $\rho_r \propto (1+z)^4$.

\subsection{Power Law Model for $F(Q)$}

We now adopt a power law form of the $f(Q)$ function due to its analytical tractability and its ability to describe deviations from General Relativity
\begin{equation}
F(Q)=6\gamma H_0^2\left(\frac{Q}{Q_0}\right)^n,
\label{eq31}
\end{equation}
where $H_0$, $\gamma$, $n$, and $Q_0$ are constants.

Using this form, the effective dark energy density and its derivative become

\begin{equation}
\rho_{de}(z) = 6\gamma H_0^2 (1-2n)
\left(\frac{H}{H_0}\right)^{2n}, \label{eq32}     
\end{equation}
and
\begin{equation}
 \dot\rho_{de} = 12n\gamma H_0^2 (1-2n)
\left(\frac{H}{H_0}\right)^{2n} \frac{\dot H}{H}. \label{eq33}   
\end{equation}

Substituting into the conservation equation yields the fundamental relation

\begin{equation}
2n \frac{\dot H}{H} + 3H(1+\omega_{de}(z)) = 0.
\label{eq34}
\end{equation}

Using the relation $\frac{d}{dt}=-(1+z)H\frac{d}{dz}$, Eq.~(\ref{eq34}) can be rewritten as

\begin{equation}
-n(1+z)\frac{d H^2}{dz}
+ 3\left[1+\omega_{de}(z)\right] H^2 = 0,
\label{eq36prime}
\end{equation}
where $\omega_{de}(z)$ corresponds to either the BA or JBP parametrization.

\subsection{Barboza-Alcaniz parametrization}

By substituting the BA parametrization, Eq.~(\ref{eq5BA}), into Eq.~(\ref{eq36prime}), we obtain
\begin{equation}
-n(1+z)\frac{dH^2}{dz}
+3\left[1+\omega_0+\omega_1
\frac{z(1+z)}{1+z^2}\right]H^2=0 .
\end{equation}
By separating the variables, we obtain
\begin{equation}
\frac{dH^2}{H^2}
=\frac{3}{n}
\left[1+\omega_0+\omega_1
\frac{z(1+z)}{1+z^2}\right]
\frac{dz}{1+z}.
\end{equation}
The general solution is given by
\begin{equation}
H^2(z)=H_0^2
(1+z)^{\frac{3(1+\omega_0)}{n}}
(1+z^2)^{\frac{3\omega_1}{2n}} .
\end{equation}

Finally, the dark energy density becomes
\begin{equation}
\rho_{\rm de}(z)=
6\gamma H_0^2(1-2n)
(1+z)^{3(1+\omega_0)}
(1+z^2)^{\frac{3\omega_1}{2}} .
\end{equation}

\subsection{Jassal-Bagla-Padmanabhan parametrization}

Following the same procedure, by substituting the JBP parametrization, Eq.~(\ref{eq5JBP}), the differential equation governing the Hubble parameter can be written as
\begin{equation}
-n(1+z)\frac{dH^2}{dz}
+3\left[1+\omega_0+\omega_1\frac{z}{(1+z)^2}\right]H^2=0,
\end{equation}

i.e.
\begin{equation}
\frac{dH^2}{H^2}
=\frac{3}{n}\left[1+\omega_0+\omega_1\frac{z}{(1+z)^2}\right]
\frac{dz}{1+z}.
\end{equation}
Therefore, the Hubble parameter evolves as
\begin{equation}
H^2(z)=H_0^2
(1+z)^{\frac{3(1+\omega_0)}{n}}
\exp\!\left[-\frac{3\omega_1}{2n}\frac{1+2z}{(1+z)^2}\right].
\end{equation}

The corresponding dark energy density becomes
\begin{equation}
\rho_{\rm de}(z)=
6\gamma H_0^2(1-2n)
(1+z)^{3(1+\omega_0)}
\exp\!\left[-\frac{3\omega_1}{2}\frac{1+2z}{(1+z)^2}\right].
\end{equation}

Hence, the normalized Hubble parameter can be written as
\begin{equation}
\begin{split}
\frac{H^2(z)}{H_0^2}=\ &\Omega_{r0}(1+z)^4+\Omega_{m0}(1+z)^3+\\
&\gamma(1-2n)(1+z)^{3(1+\omega_0)}f_{\rm de}(z),
\end{split}
\end{equation}
where
\begin{equation}
f_{\rm de}(z)=
\begin{cases}
(1+z^2)^{\frac{3\omega_1}{2}}, & \text{BA parametrization},\\[0.6em]
\exp\!\left[-\frac{3\omega_1}{2}\frac{1+2z}{(1+z)^2}\right], & \text{JBP parametrization}.
\end{cases}
\end{equation}

Once the Hubble function is determined, the background evolution of the Universe follows in a straightforward way. Matter and radiation evolve according to their usual scaling laws, while the effective dark energy density inherits the same logarithmic behavior encoded in the equation of state. The resulting Friedmann equation provides the starting point for the observational analysis presented in the next section.

For consistency with standard cosmological practice, we adopt the normalization
\begin{equation}
H_0 = 100\,h \;\mathrm{km\,s^{-1}\,Mpc^{-1}}.
\end{equation}

Within this framework, the parameter $\gamma$ is not independent but is instead determined by the present-day matter density $\Omega_{m0}$ and the power law index $n$ through
\begin{equation}
\gamma(\Omega_{m0}, n)= \frac{1 - \Omega_{m0} - \Omega_{r0}}{1 - 2 n},
\end{equation}
where $\Omega_{r0}$ is the current radiation density parameter, fixed at $\Omega_{r0} = 8.4 \times 10^{-5}$. This ensures that the model is properly normalized at the present epoch.

\section{Observational Data and Statistical Methodology}
\label{sec:3}

To constrain the parameters of the power law $f(Q)$ model, we employ a Markov Chain Monte Carlo (MCMC) technique, which allows for an efficient exploration of the parameter space and a reliable estimation of posterior distributions~\cite{refA}. The analysis is performed by combining multiple cosmological observations, ensuring robust constraints through the complementarity of the datasets.
The resulting MCMC chains are then used to extract the best fit values of the model parameters, along with their corresponding confidence~\cite{refB,refC}.

In our analysis, we constrain the model parameters using four independent observational datasets: the Pantheon$^{+}$ Type Ia supernova compilation, consisting of 1701 data points~\cite{refD}; the DESI DR2 BAO measurements, comprising 12 data points~\cite{refE}; the Cosmic Chronometer (CC) observations, including 32 data points~\cite{refF}; and the Redshift Space Distortion (RSD) dataset, with 20 data points~\cite{Dalale1}. This combination enables us to probe both the expansion and the growth history of the Universe over a wide redshift range, while significantly reducing parameter degeneracies and improving the robustness of the constraints.

\subsection{Pantheon$^{+}$ dataset}

We use the Pantheon$^{+}$ compilation of Type Ia supernovae~\cite{refD}, which consists of 1701 measurements from 1550 supernovae spanning the redshift range $0.001 \leq z \leq 2.3$. The corresponding likelihood is constructed through the chi-square function
\begin{equation}
\chi_{\text{Pantheon}^{+}}^2
= \vec{F}^{\,T} \cdot \mathbf{C}_{\text{Pantheon}^{+}}^{-1} \cdot \vec{F},
\label{eq11}
\end{equation}
where $\mathbf{C}_{\text{Pantheon}^{+}}$ is the full covariance matrix, including both statistical and systematic uncertainties. The data vector is defined as
\begin{equation}
\vec{F}_i = m_{B,i} - M - \mu_{\rm model},
\end{equation}
where $m_{B,i}$ and $\mu_{\rm model}$ denote the observed apparent magnitude and the theoretical distance modulus, respectively.

The theoretical distance modulus is given by
\begin{equation}
\mu_{\rm model}(z) = 5 \log_{10} D_L(z) + 25,
\end{equation}
with the luminosity distance defined as
\begin{equation}
D_L(z) = (1+z)\int_0^z \frac{dz^\ast}{H(z^\ast)}.
\end{equation}

Unlike the original Pantheon sample, Pantheon$^{+}$ partially breaks the degeneracy between the absolute magnitude $M$ and the Hubble constant $H_0$ by incorporating supernovae hosted in Cepheid-calibrated galaxies. This leads to a modified data vector,
\begin{equation}
\vec{F}_i^{\prime} =
\begin{cases}
m_{B,i} - M - \mu_i^{\rm Ceph}, & i \in \text{Cepheid hosts}, \\[0.8em]
m_{B,i} - M - \mu_{\rm model}(z_i), & \text{otherwise},
\end{cases}
\end{equation}
where $\mu_i^{\rm Ceph}$ is the independently calibrated distance modulus from Cepheid variables.

Consequently, the chi-square function becomes
\begin{equation}
\chi_{\mathrm{Pantheon}^{+}}^2
= \vec{F}^{\prime \, T} \cdot \mathbf{C}_{\text{Pantheon}^{+}}^{-1} \cdot \vec{F}^{\prime}.
\end{equation}

\subsection{DESI DR2 dataset}

We use the DESI DR2 BAO dataset, which includes multiple tracers such as Bright Galaxy Survey (BGS) galaxies, Luminous Red Galaxies (LRGs), Emission Line Galaxies (ELGs), quasars (QSO), and the Lyman-$\alpha$ forest, covering the redshift range $0.1 \leq z \leq 4.2$ ~\cite{refE}. 

From these observations, we consider the comoving angular diameter distance $D_M(z)/r_d$ and the Hubble distance $D_H(z)/r_d$, defined as
\begin{equation}
D_M(z) = \int_0^z \frac{c \, dz'}{H(z')}, \quad
D_H(z) = \frac{c}{H(z)},
\end{equation}
where $r_d$ denotes the sound horizon at the drag epoch. Assuming standard early Universe physics, it is given by
\begin{equation}
r_d = \int_{z_d}^{\infty} \frac{c_s(z')}{H(z')} \, dz',
\end{equation}
with $z_d$ the drag epoch redshift and $c_s$ the sound speed of the baryon photon fluid. We also include the volume-averaged distance
\begin{equation}
D_V(z) = \left[z \, D_M^2(z) \, D_H(z)\right]^{1/3}.
\end{equation}

The DESI DR2 measurements used in this work are taken from Table~III of~\cite{refG}, together with their corresponding covariance matrices, $\mathbf{C_{DESI}}$. The associated chi-square function is defined as
\begin{equation}
\chi^2_{\mathrm{DESI}} =
\Delta D^{T}\, \mathbf{C_{DESI}}^{-1}\, \Delta D,
\end{equation}
where $\Delta D = D^{\mathrm{obs}}_i - D^{\mathrm{th}}_i$ represents the difference between the observed and theoretical BAO measurements, and $\mathbf{C_{DESI}}$ denotes the covariance matrix of the dataset. In the case of uncorrelated data, the chi-square simplifies to
\begin{equation}
\chi^2_{\mathrm{DESI}} = \sum_i \left(\frac{D^{\mathrm{obs}}_i - D^{\mathrm{th}}_i}{\sigma_i}\right)^2.
\end{equation}

\subsection{Cosmic Chronometers}

In addition, we use 32 cosmic chronometer (CC) measurements of the Hubble parameter $H(z)$ in the redshift range $0 < z \lesssim 2$~\cite{refF}. The CC method provides a direct and nearly model independent estimate of the expansion rate, based on the differential age evolution of passively evolving galaxies, through
\begin{equation}
H(z) = -\frac{1}{1+z} \frac{dz}{dt}.
\end{equation}

The dataset includes a full covariance matrix to account for correlated systematic uncertainties, such as those related to metallicity, stellar population synthesis models, and contamination from young stellar populations. The chi-square function is defined as
\begin{equation}
\chi^2_{\mathrm{CC}} = \Delta \mathbf{H}^T \, \mathbf{C_{CC}}^{-1} \, \Delta \mathbf{H},
\end{equation}
where $\Delta \mathbf{H} = H_{\mathrm{th}}(z_i) - H_{\mathrm{obs}}(z_i)$, with $H_{\mathrm{obs}}(z_i)$ and $H_{\mathrm{th}}(z_i)$ representing the observed and theoretical values of the Hubble parameter at redshift $z_i$, respectively. The covariance matrix $\mathbf{C_{CC}}$ accounts for both statistical and systematic uncertainties.

\subsection{Redshift Space Distortion dataset}
\label{rsd}

In order to further constrain the growth of cosmic structures, we incorporate measurements of redshift space distortions (RSD), which provide direct information on the growth rate of matter perturbations. In particular, we consider observational data of the quantity $f\sigma_8(z)$, defined as the product of the growth rate $f(z)$ and the amplitude of matter fluctuations at the scale $8h^{-1}\mathrm{Mpc}$.

We use a compilation of 20 data points~\cite{Dalale1}, spanning the redshift range $0.02 \leq z \leq 1.944$. The corresponding chi-square function is defined as
\begin{equation}
\chi^2_{\mathrm{RSD}} = \sum_{i=1}^{20} \left( \frac{f\sigma_{8,\mathrm{obs}}(z_i) - f\sigma_{8,\mathrm{th}}(z_i)}{\sigma(z_i)} \right)^2 .
\label{eq:chi2_rsd}
\end{equation}
where $\sigma(z_i)$ denotes the observational uncertainty associated with each data point.

The theoretical prediction for $f\sigma_8$ is given by
\begin{equation}
f\sigma_{8,\mathrm{th}}(z_i) = \sigma_8 \frac{\delta'(z_i)}{\delta(z=0)},
\label{eq:fs8_th}
\end{equation}
where the prime denotes differentiation with respect to $x = \ln(a)$, and $\sigma_8$ is the present day amplitude of the matter power spectrum.

To evaluate this quantity, we numerically solve the evolution equation of the matter density contrast in the quasi-static regime
\begin{equation}
\ddot{\delta}_m + 2H\dot{\delta}_m - 4\pi G_{\mathrm{eff}} \rho_m \delta_m = 0,
\label{eq:delta_t}
\end{equation}
where the dot represents differentiation with respect to cosmic time, and $G_{\mathrm{eff}} $ is the effective gravitational coupling. This gravitational coupling constitutes the main difference between modified gravity theories and general relativity, particularly at the perturbative level, as shown in Eq.~(\ref{eq:delta_t}). This effective gravitational coupling is  defined as 
\begin{equation}
G_{\mathrm{eff}} = \frac{G}{f_Q} = G \left[
1+\gamma\, n\, E^{2(n-1)}(z)
\right]^{-1},
\label{eq417}
\end{equation}
where $E(z)=H(z)/H_0$ is the normalized Hubble parameter. Such an effective gravitational coupling  encodes deviations from standard gravity and characterizes their impact on the growth of matter perturbations and structure formation~\cite{Geff}. For numerical convenience, this equation can be rewritten as
\begin{equation}
\delta_m'' + \left( \frac{H'(x)}{H(x)} + 2 \right)\delta_m' - \frac{4\pi G\rho_m}{(1 + F_Q)H(x)^2} \delta_m = 0 .
\label{eq:delta_x}
\end{equation}

Assuming that the different datasets are statistically independent, the total chi-square is given by~\cite{refA}
\begin{equation}
\chi^2_{\mathrm{tot}} =
\chi^2_{\mathrm{DESI}} +
\chi^2_{\mathrm{SN}} +
\chi^2_{\mathrm{CC}} +
\chi^2_{\mathrm{RSD}}.
\end{equation}

This combined analysis provides robust and tightly constrained cosmological parameters by exploiting the complementarity between geometric (Pantheon$^{+}$, BAO), expansion-history (CC), and growth-of-structure (RSD) probes.

Finally, we perform the analysis using three dataset combinations: Pantheon$^{+}$+DESI DR2, Pantheon$^{+}$+DESI DR2+CC, and Pantheon$^{+}$+DESI DR2+CC+RSD with the adopted priors summarized in Table~\ref{tab:1}.

By evaluating the minimum chi-square value, $\chi^2_{\min}$, we derive statistical criteria to compare models with different numbers of free parameters. In this work, we adopt the corrected Akaike Information Criterion (AIC$_c$) and the Bayesian Information Criterion (BIC), defined respectively as
\begin{equation}
\mathrm{AIC_c} = -2 \ln \mathcal{L}_{\max} + 2 N_p + \frac{2 N_p (N_p + 1)}{N_d - N_p - 1},
\end{equation}
and
\begin{equation}
\mathrm{BIC} = -2 \ln \mathcal{L}_{\max} + N_p \ln (N_d),
\end{equation}
where $N_d$ and $N_p$ denote the number of data points and free parameters, respectively.

The statistical performance of each model is evaluated relative to the $\Lambda$CDM scenario through
\begin{equation}
\Delta\mathrm{AIC} = \mathrm{AIC}_{\text{model}} - \mathrm{AIC}_{\Lambda\mathrm{CDM}},
\end{equation}
\begin{equation}
\Delta\mathrm{BIC} = \mathrm{BIC}_{\text{model}} - \mathrm{BIC}_{\Lambda\mathrm{CDM}}.   
\end{equation}

The interpretation follows standard criteria: $|\Delta \mathrm{AIC}_c| < 2$ indicates statistically indistinguishable models, $2 < |\Delta \mathrm{AIC}_c| < 4$ suggests moderate evidence against the model with the larger value, $4 < |\Delta \mathrm{AIC}_c| < 6$ indicates positive evidence, while $|\Delta \mathrm{AIC}_c| > 6$ corresponds to strong to decisive evidence against it. A similar interpretation applies to the BIC criterion.

\section{Results and discussion}
\label{sec:4}

In Table~\ref{tab:1}, we present the mean values of the model parameters along with their $1\sigma$ uncertainties for the three cosmological models, namely $\Lambda$CDM, BA, and JBP, using the Pantheon$^{+}$+DESI DR2, Pantheon$^{+}$+DESI DR2+CC, and Pantheon$^{+}$+DESI DR2+CC+RSD datasets. The $\Lambda$CDM model is described by the parameter set $(\Omega_m, h, M, r_d, \sigma_8)$, while the BA and JBP parametrizations extend this set to $(\Omega_m, h, M, \omega_0, \omega_1, n, \gamma, r_d, \sigma_8)$.

The $\Lambda$CDM model exhibits remarkably stable constraints across all dataset combinations. For Pantheon$^{+}$+DESI DR2, we obtain $\Omega_m = 0.3070 \pm 0.0097$ and $h = 0.736 \pm 0.011$, while the inclusion of CC data yields $\Omega_m = 0.3075 \pm 0.0099$ and $h = 0.735 \pm 0.011$. When RSD data are added, the background parameters remain essentially unchanged; however, the growth sector becomes constrained through $\sigma_8 = 0.791 \pm 0.024$, highlighting the role of RSD measurements in probing the matter perturbation amplitude. The parameters $M$ and $r_d$ also remain consistent across all cases.

For the BA parametrization, the Pantheon$^{+}$+DESI DR2 dataset gives $\Omega_m = 0.271^{+0.020}_{-0.018}$ and $h = 0.733 \pm 0.011$, with $\omega_0 = -0.890 \pm 0.052$ and $\omega_1 = 0.130 \pm 0.046$. Including CC data shifts the constraints to $\Omega_m = 0.292 \pm 0.015$ and $h = 0.732 \pm 0.010$, with $\omega_0 = -0.895 \pm 0.046$ and $\omega_1 = -0.012^{+0.097}_{-0.076}$. The additional parameters are constrained to $n = -1.44^{+0.14}_{-0.19}$ and $\gamma = 0.183^{+0.014}_{-0.021}$.In this case, the amplitude of matter fluctuations is constrained to $\sigma_8 = 0.819 \pm 0.031$, which deviates from the $\Lambda$CDM by approximately $0.7\sigma$.

Similarly, for the JBP parametrization, the Pantheon$^{+}$+DESI DR2 dataset yields $\Omega_m = 0.259^{+0.028}_{-0.024}$ and $h = 0.775^{+0.030}_{-0.041}$, with $\omega_0 = -0.900 \pm 0.049$ and $\omega_1 = 0.098^{+0.070}_{-0.089}$. Adding CC data leads to $\Omega_m = 0.261^{+0.022}_{-0.019}$ and $h = 0.734 \pm 0.011$, with $\omega_0 = -0.889 \pm 0.052$ and $\omega_1 = 0.267 \pm 0.051$, while $n = -1.05^{+0.17}_{-0.12}$ and $\gamma = 0.240^{+0.025}_{-0.020}$. When RSD data are included, the constraints become $\Omega_m = 0.314^{+0.029}_{-0.035}$ and $h = 0.710^{+0.035}_{-0.030}$, with the corresponding value of $\sigma_8$ constrained to $0.810 \pm 0.030$, deviating from the $\Lambda$CDM prediction by approximately $0.5\sigma$. This clearly shows that the inclusion of RSD data directly impacts the perturbation sector while leaving the background evolution largely consistent.

Overall, the addition of CC and RSD datasets improves the robustness of the constraints. In particular, RSD measurements play a crucial role in breaking degeneracies and providing direct information on the growth of matter perturbations through $\sigma_8$, complementing the background constraints from SNIa, BAO, and CC data.

In Table~\ref{tab:2}, we present the values of $\chi^2_{\min}$, $\mathrm{AIC}_c$, and BIC for the different models. The BA and JBP parametrizations yield lower minimum $\chi^2$ values than $\Lambda$CDM for both dataset combinations. However, since these models contain a larger number of free parameters, the reduction in $\chi^2_{\min}$ alone does not provide sufficient evidence for a better model. Therefore, we further assess the models using the $\mathrm{AIC}_c$ and $\mathrm{BIC}$ criteria, which account for model complexity. The $\mathrm{AIC}_c$ differences remain small, with $\Delta \mathrm{AIC}_c$(BA) $\approx 1.23$ and $\Delta \mathrm{AIC}_c$(JBP) $\approx 1.54$ for Pantheon$^{+}$+DESI DR2+CC, and $\Delta \mathrm{AIC}_c$(BA) $\approx 2.15$ and $\Delta \mathrm{AIC}_c$(JBP) $\approx 2.10$ when including RSD, indicating that BA and JBP remain competitive with \(\Lambda\)CDM, with no substantial evidence favoring one model over another. In contrast, the BIC strongly disfavors these models, with $\Delta \mathrm{BIC}$(BA) $\approx 17.58$ and $\Delta \mathrm{BIC}$(JBP) $\approx 17.89$ for Pantheon$^{+}$+DESI DR2+CC, and $\Delta \mathrm{BIC}$(BA) $\approx 18.52$ and $\Delta \mathrm{BIC}$(JBP) $\approx 18.47$ when including RSD, reflecting the stronger penalty imposed by the BIC for their larger number of free parameters.

\begin{figure}[H]
    \centering
    \includegraphics[width=1.0\linewidth]{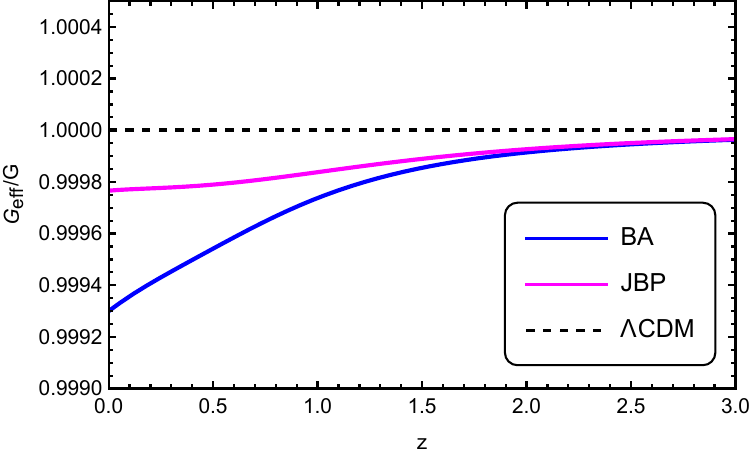}
    \caption{The evolution of the effective Newton’s constant normalized by the gravitational constant, $G_{\mathrm{eff}}/G$, as a function of redshift.}
    \label{figGeff}
\end{figure}

As shown in Fig.~\ref{figGeff}, both BA and JBP parametrizations yield $G_{\mathrm{eff}}/G < 1$ over the full redshift range, with the deviation from unity being largest at low redshift ($z \approx 0$) and gradually vanishing as z increases, approaching the $\Lambda$CDM value ($G_{\mathrm{eff}}/G = 1$) at high redshift. This behavior indicates a slightly weaker effective gravitational interaction compared to General Relativity, which may lead to a mild suppression of matter clustering. At the same time, the small amplitude of the deviation shows that the models remain close to the standard cosmological scenario and are consistent with current structure-growth observations.

\begin{figure*}
    \centering
    \includegraphics[width=0.99\linewidth]{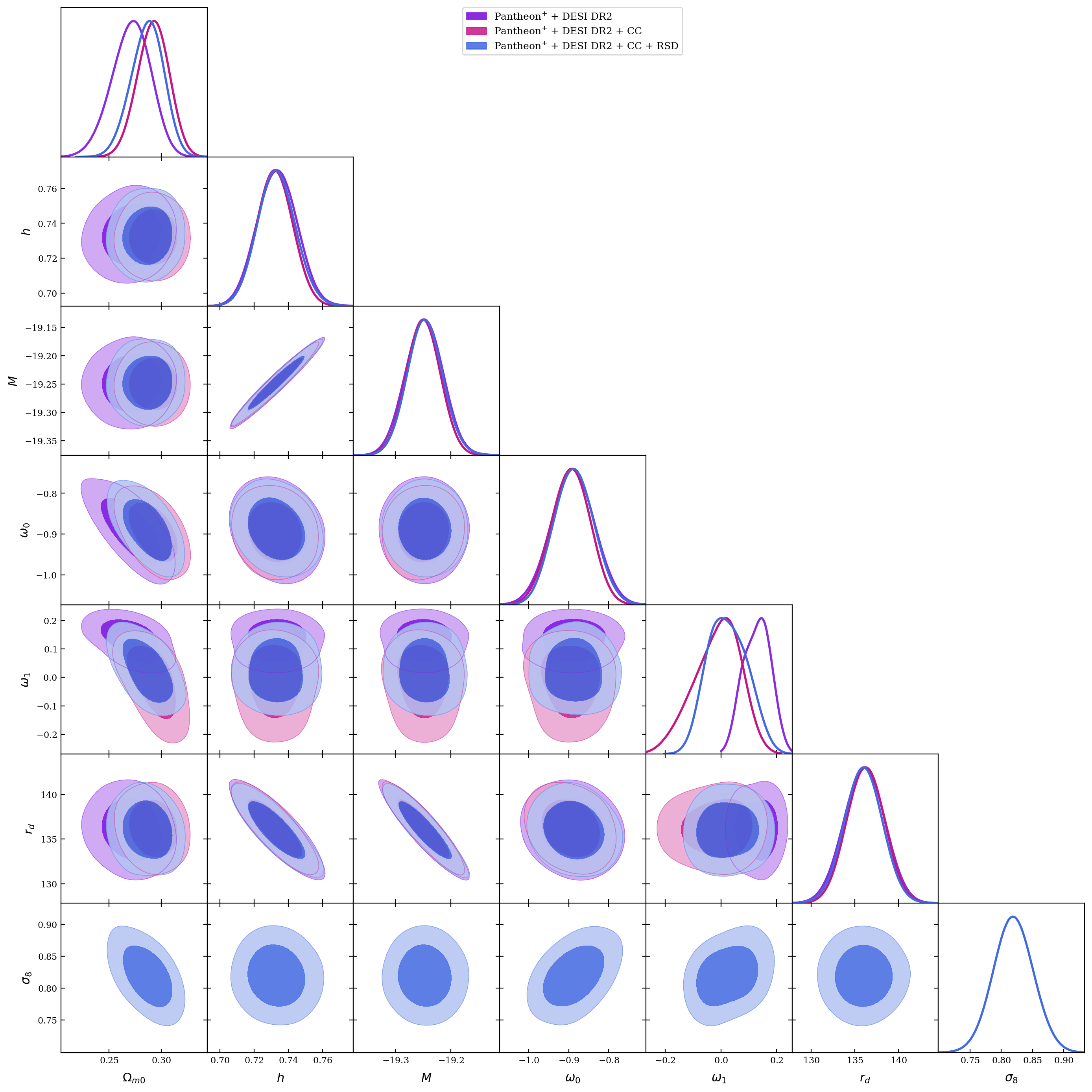}
    \caption{The $1\sigma$ and $2\sigma$ confidence contours and posterior distributions for the BA model using the combined Pantheon$^{+}$, DESI DR2, CC, and RSD datasets.}
    \label{fig:1}
\end{figure*}

\begin{figure*}
    \centering
    \includegraphics[width=0.99\linewidth]{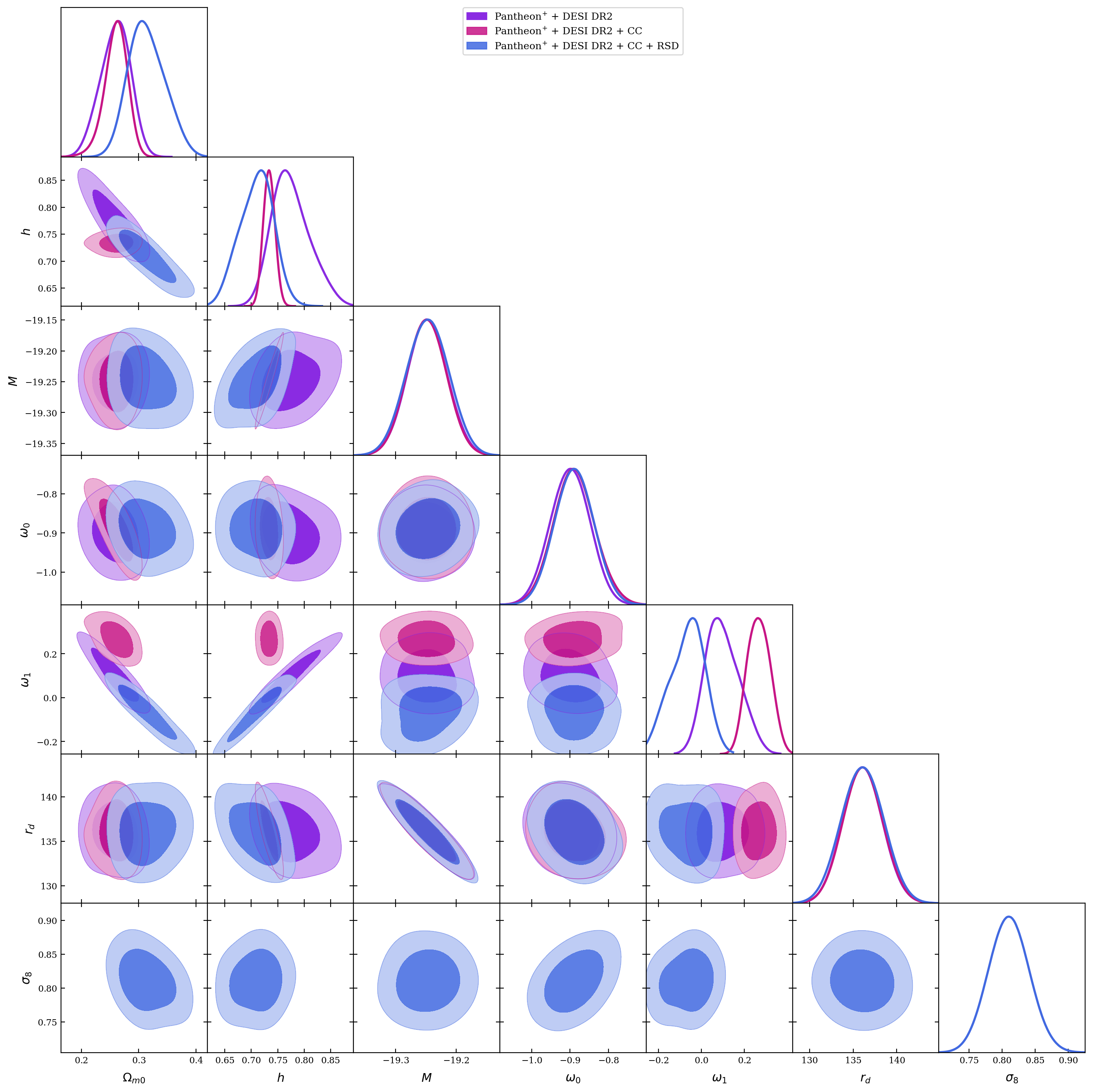}
    \caption{The $1\sigma$ and $2\sigma$ confidence contours and posterior distributions for the JBP model using the combined Pantheon$^{+}$, DESI DR2, CC, and RSD datasets.}
    \label{fig:2}
\end{figure*}

\begin{figure*}
    \centering
    \includegraphics[width=0.8\linewidth]{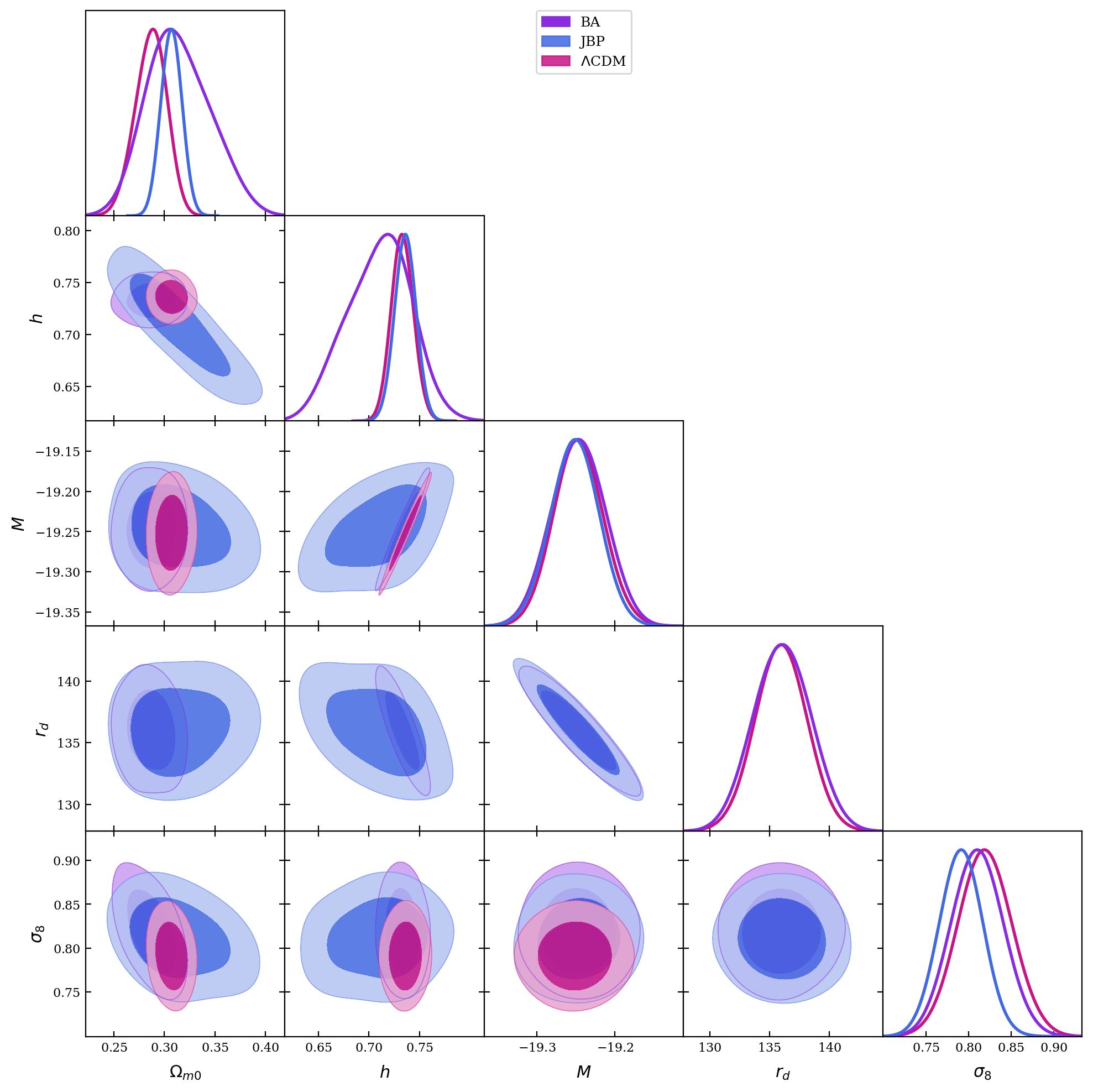}
    \caption{The $1\sigma$ and $2\sigma$ confidence contours, along with the posterior distributions, obtained for the BA, JBP, and $\Lambda$CDM models using the combined Pantheon$^{+}$, DESI DR2, CC, and RSD datasets.}
    \label{fig:3}
\end{figure*}

Additionally, Figs.~\ref{fig:1} and \ref{fig:2} present the $1\sigma$ and $2\sigma$ confidence contours for the BA and JBP models, respectively, obtained from the combined datasets. These plots also illustrate the correlations among the model parameters at both confidence levels. In particular, a positive correlation is observed between $M$, $\omega_0$, and $r_d$, which remain stable across all datasets, while a negative correlation is found between $\Omega_{m0}$ and $\omega_1$. The inclusion of RSD data significantly tightens the constraints, particularly on $\sigma_8$. Overall, adding CC and RSD measurements improves the parameter constraints and reduces the allowed parameter space for both models. Furthermore, Fig.~\ref{fig:3} highlights the comparison between the BA, JBP, and $\Lambda$CDM models, showing that both BA and JBP exhibit slight deviations from the $\Lambda$CDM contours while remaining statistically consistent within the $1\sigma$ and $2\sigma$ confidence levels.

\begin{table*}[t]
\centering
\caption{Mean values of the parameters for the different cosmological models using the Pantheon$^+$+DESI DR2, Pantheon$^+$+DESI DR2+CC, and Pantheon$^+$+DESI DR2+CC+RSD datasets.}
\label{tab:1}
\begin{tabular}{l | l | l | c | c | c}
\hline
Models & Parameters & Priors & Pantheon$^+$+DESI DR2 & Pantheon$^+$+DESI DR2+CC & Pantheon$^+$+DESI DR2+CC+RSD \\
\hline\hline

\multirow{4}{*}{$\Lambda$CDM} 
& $\Omega_m$ & [0,1] 
& $0.3070 \pm 0.0097$ 
& $0.3075 \pm 0.0099$
& $0.3069 \pm 0.0098$\\

& $h$ & [0.4,2]  
& $0.736 \pm 0.011$ 
& $0.735 \pm 0.011$
& $0.736 \pm 0.010$\\

& $M$ & [-20,-19] 
& $-19.252 \pm 0.031$ 
& $-19.253 \pm 0.031$
& $-19.251 \pm 0.029$\\

& $r_d$ & [100,200] 
& $137.0 \pm 2.1$ 
& $137.1 \pm 2.1$
& $137.0 \pm 2.1$\\

& $\sigma_8$ & [0,2] 
& -- 
& --
& $0.791 \pm 0.024$ \\

\hline

\multirow{8}{*}{BA}
& $\Omega_m$ & [0,1] 
& $0.271^{+0.020}_{-0.018}$ 
& $0.292 \pm 0.015$
& $0.287^{+0.017}_{-0.015}$ \\

& $h$ & [0.4,2] 
& $0.733 \pm 0.011$ 
& $0.732 \pm 0.010$ 
& $0.733 \pm 0.011$ \\

& $M$ & [-20,-19] 
& $-19.248 \pm 0.032$   
& $-19.250 \pm 0.029$
& $-19.247 \pm 0.031$\\

& $\omega_0$ & [-2,2] 
& $-0.890 \pm 0.052$ 
& $-0.895 \pm 0.046$
& $-0.887 \pm 0.048$\\

& $\omega_1$ & [-2,2] 
& $0.130 \pm 0.046$ 
& $-0.012^{+0.097}_{-0.076}$ 
& $0.023^{+0.069}_{-0.078}$ \\

& $n$ & [-2,2] 
& $-1.044^{+0.013}_{-0.015}$ 
& $-1.44^{+0.14}_{-0.19}$
& $-1.110 \pm 0.063$\\

& $\gamma$ & 
& $0.236 \pm 0.0055$ 
& $0.183^{+0.014}_{-0.021}$ 
& $0.222^{+0.011}_{-0.014}$ \\

& $r_d$ & [100,200] 
& $136.1 \pm 2.2$ 
& $136.2 \pm 2.1$
& $136.0 \pm 2.1$\\

& $\sigma_8$ & [0,2] 
& -- 
& --
& $0.819 \pm 0.031$\\

\hline

\multirow{8}{*}{JBP}
& $\Omega_m$ & [0,1] 
& $0.259^{+0.028}_{-0.024}$ 
& $0.261^{+0.022}_{-0.019}$ 
& $0.314^{+0.029}_{-0.035}$ \\

& $h$ & [0.4,2] 
& $0.775^{+0.030}_{-0.041}$ 
& $0.734 \pm 0.011$
& $0.710^{+0.035}_{-0.030}$\\

& $M$ & [-20,-19] 
& $-19.248 \pm 0.031$                          & $-19.248 \pm 0.031$
& $-19.247 \pm 0.033$\\

& $\omega_0$ & [-2,2] 
& $-0.900 \pm 0.049$ 
& $-0.889 \pm 0.052$
& $-0.889 \pm 0.049$\\

& $\omega_1$ & [-2,2] 
& $0.098^{+0.070}_{-0.089}$ 
& $0.267 \pm 0.051$
& $-0.065^{+0.091}_{-0.072}$\\

& $n$ & [-2,2] 
& $-0.991^{+0.025}_{-0.018}$ 
& $-1.05^{+0.17}_{-0.12}$
& $-1.02^{+0.11}_{-0.13}$\\

& $\gamma$ & 
& $0.249 \pm 0.011$ 
& $0.240^{+0.025}_{-0.020}$
& $0.228^{+0.020}_{-0.027}$\\

& $r_d$ & [100,200] 
& $136.1 \pm 2.2$ 
& $136.1 \pm 2.2$
& $136.1 \pm 2.3$\\

& $\sigma_8$ & [0,2] 
& -- 
& --
& $0.810 \pm 0.030$\\

\hline
\end{tabular}
\end{table*}

\begin{table*}[t]
\centering
\caption{The corresponding $\chi^2_{\min}$, $\mathrm{AIC}_c$, and BIC values for the examined cosmological models using the Pantheon$^+$+DESI DR2+CC and Pantheon$^+$+DESI DR2+CC+RSD dataset combinations.}
\label{tab:2}
\begin{tabular}{l|c|c|c|c|c}
\hline\hline
Models & $\chi^2_{\min}$ & AIC$_c$ & $\Delta$AIC$_c$ & BIC & $\Delta$BIC \\
\hline\hline
\multicolumn{6}{c}{Pantheon$^+$ + DESI DR2 + CC} \\
\hline
$\Lambda$CDM & 1546.71 & 1554.73 & 0 & 1576.56 & 0  \\
BA & 1541.90 & 1555.96 & 1.23 & 1594.14 & 17.58 \\
JBP & 1542.21 & 1556.27 & 1.54 & 1594.45 & 17.89 \\
\hline\hline
\multicolumn{6}{c}{Pantheon$^+$ + DESI DR2 + CC + RSD} \\
\hline
$\Lambda$CDM & 1554.80 & 1564.83 & 0 & 1592.18 & 0 \\
BA & 1550.90 & 1566.98 & 2.15 & 1610.70 & 18.52 \\
JBP & 1550.85 & 1566.93 & 2.10 & 1610.65 & 18.47 \\
\hline
\end{tabular}
\end{table*}

\section{Dynamical analysis}
\label{sec:5}

\subsection{Cosmographic parameters}

To discriminate between the numerous cosmological models proposed to explain the late-time acceleration of the Universe, it is useful to adopt a model independent approach. Cosmography provides such a framework, as it relies only on the kinematics of the cosmic expansion, without assuming any specific dark energy scenario \cite{D1,D2,D3}. 

This method is based on the Taylor expansion of the scale factor $a(t)$ around the present time, allowing cosmological models to be characterized through its successive time derivatives \cite{D4,D5}. The main cosmographic parameters are defined as
\begin{equation}
H(t) = \frac{1}{a} \frac{d a}{d t},
\end{equation}

\begin{equation}
q(t) = -\frac{1}{a H^2} \frac{d^2 a}{d t^2},
\end{equation}

\begin{equation}
j(t) = \frac{1}{a H^3} \frac{d^3 a}{d t^3},
\end{equation}

\begin{equation}
s(t) = \frac{1}{a H^4} \frac{d^4 a}{d t^4}.
\end{equation}

In terms of the redshift $z$, these quantities can be expressed as
\begin{equation}
q(z) = -1 + (1+z)\frac{E'(z)}{E(z)},
\end{equation}
\begin{equation}
j(z) = (1+z)^2 \frac{E''(z)}{E(z)} + q^2(z),
\end{equation}
and
\begin{equation}
s(z) = -(1+z) j'(z) - 2 j(z) - 3 q(z) j(z).
\end{equation}

\begin{figure*}
    \centering
     \includegraphics[width=0.5\linewidth]{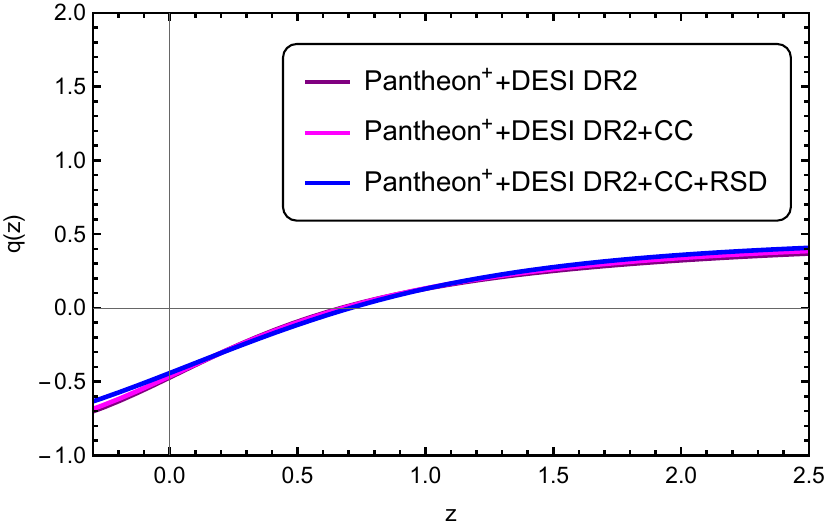}\includegraphics[width=0.5\linewidth]{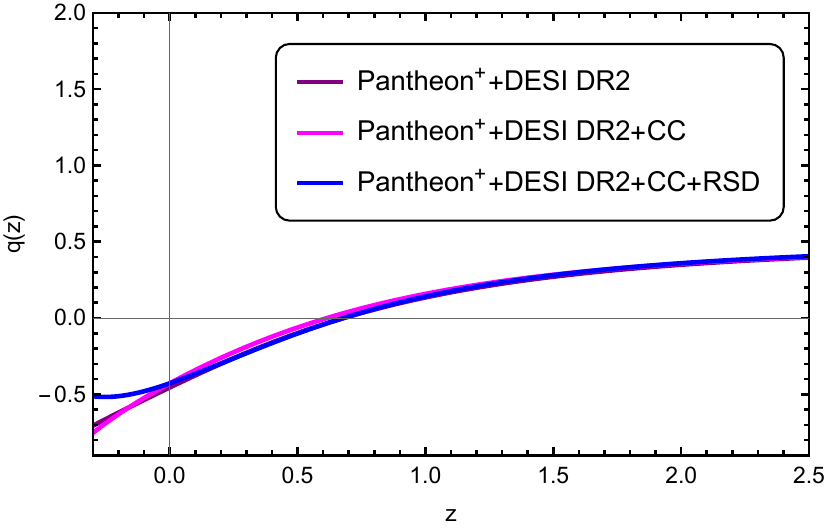}
    \caption{The evolution of the deceleration parameter $q(z)$ for the BA model (left panel) and the JBP model (right panel) as functions of redshift.}
    \label{fig:6}
\end{figure*}

\begin{figure*}
    \centering
    \includegraphics[width=0.5\linewidth]{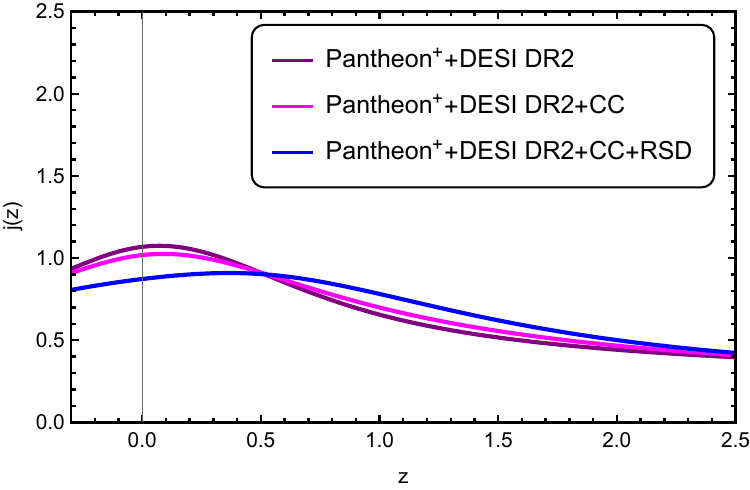}\includegraphics[width=0.5\linewidth]{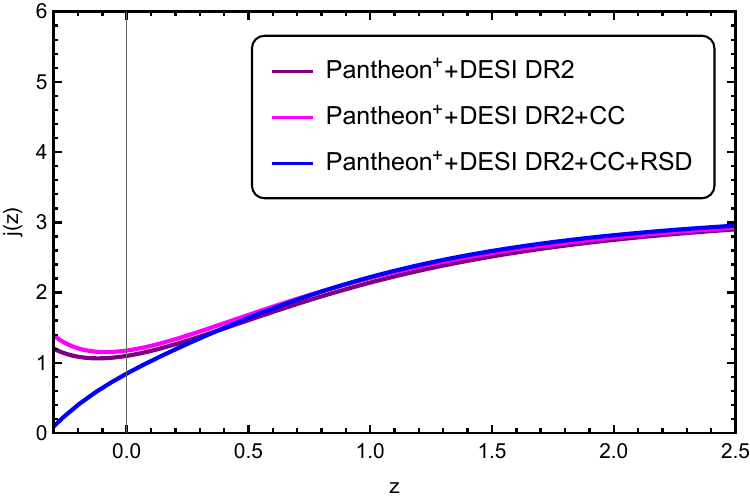}
    \caption{The evolution of the jerk parameter $j(z)$ for the BA model (left panel) and the JBP model (right panel) as functions of redshift.}
    \label{fig:7}
\end{figure*}

\begin{figure*}
    \centering
    \includegraphics[width=0.5\linewidth]{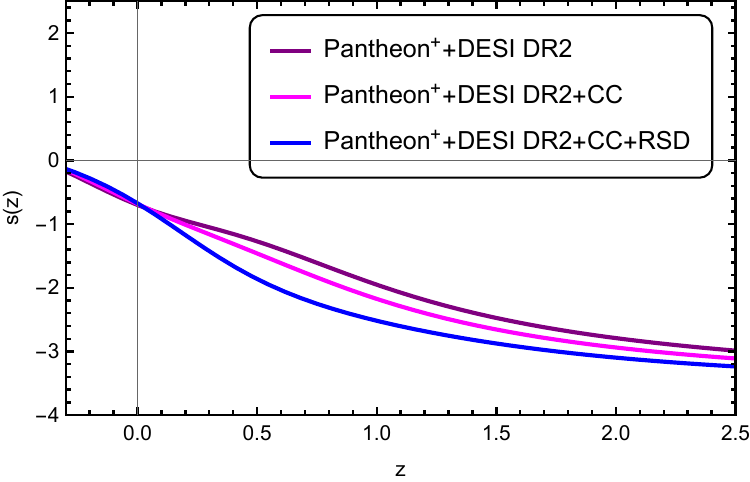}\includegraphics[width=0.5\linewidth]{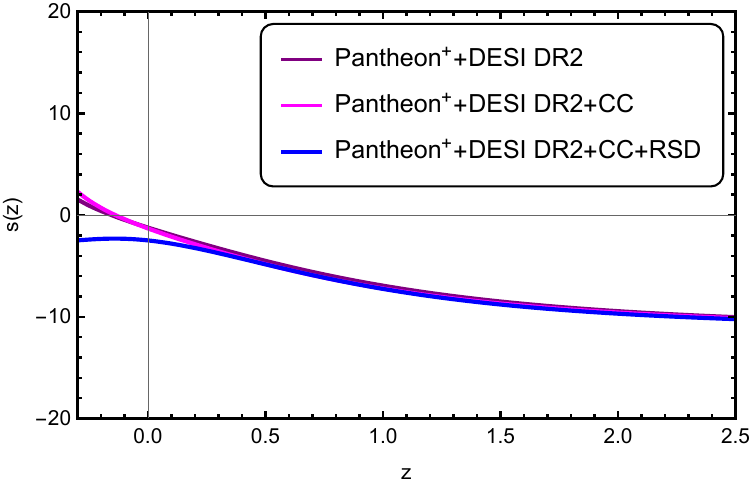}
    \caption{The evolution of the snap parameter $s(z)$ for the BA model (left panel) and the JBP model (right panel) as functions of redshift.}
    \label{fig:9}
\end{figure*}

Figs.~\ref{fig:6}, \ref{fig:7}, and \ref{fig:9} collectively illustrate the redshift evolution of the cosmographic parameters $q(z)$, $j(z)$, and $s(z)$ for both the BA and JBP parametrizations, using the Pantheon$^{+}$+DESI DR2, Pantheon$^{+}$+DESI DR2+CC datasets, and Pantheon$^{+}$+DESI DR2+CC+RSD. For the Pantheon$^{+}$+DESI DR2+CC+RSD dataset, a similar smooth and well-behaved evolution is observed in all three parameters, with no indication of divergences or pathological features. Together, these parameters provide a comprehensive kinematical description of the cosmic expansion up to higher orders. In all cases, the reconstructed functions exhibit smooth and well behaved evolutions over the entire redshift range.

The deceleration parameter $q(z)$, shown in Fig.~\ref{fig:6}, exhibits a clear transition from an early matter dominated decelerating phase ($q>0$) to the present accelerated expansion regime ($q<0$), in agreement with the standard cosmological scenario. For the BA parametrization, this transition occurs at intermediate redshifts, with $z_t \approx 0.672$ and $z_t \approx 0.679$ depending on whether CC data are included. When the full Pantheon$^{+}$+DESI DR2+CC+RSD dataset is considered, the transition redshift is found to be $z_t = 0.708$. This indicates that the inclusion of RSD data does not significantly shift the transition redshift, suggesting that the model remains stable under the addition of growth-rate measurements. At the present epoch, the deceleration parameter takes negative values, with $q_0 \approx -0.472$ and $q_0 \approx -0.464$, while the full dataset yields $q_0 = -0.442$, consistently confirming the current accelerated expansion of the Universe. The close agreement between these values highlights the stability of the BA parametrization under the inclusion of additional observational data. For the JBP model, a similar dynamical behavior is observed. The transition redshift is found to be $z_t \approx 0.680$ and $z_t \approx 0.535$, depending on the inclusion of CC data, while the inclusion of RSD data gives $z_t = 0.678$, showing a consistent evolution pattern comparable to the BA case. At the present epoch, the deceleration parameter remains negative, with $q_0 \approx -0.455$ and $q_0 \approx -0.394$, and for the full Pantheon$^{+}$+DESI DR2+CC+RSD dataset we obtain $q_0 = -0.428$, further confirming the robustness of the late-time accelerated expansion across both parametrizations.

The jerk parameter $j(z)$, shown in Fig.~\ref{fig:7}, remains close to the $\Lambda$CDM expectation ($j=1$) for the BA parametrization, with present day values $j_0 = 0.81$ and $0.77$, indicating only mild deviations from the standard cosmological model and a weak sensitivity to the inclusion of CC data. For the Pantheon$^{+}$+DESI DR2+CC+RSD dataset, we obtain $j_0 = 0.87$ (BA). In contrast, the JBP parametrization exhibits a slightly stronger redshift dependence, with $j_0 = 1.1$ and $1.3$, while for the full dataset we find $j_0 = 0.85$, indicating that the inclusion of RSD data reduces the deviation from the $\Lambda$CDM value and improves consistency with the standard cosmological scenario.

The snap parameter $s(z)$ further characterizes the fine structure of the expansion dynamics. For the BA model, $s(z)$ shows a gradual evolution toward negative values, with $s_0 = -0.70$ and $-0.68$, again indicating only minor shifts induced by the inclusion of CC data. This behavior remains consistent when RSD data are included, with $s_0 = -0.67$. Conversely, the JBP parametrization displays a steeper behavior, reaching more negative values, with $s_0 = -1.20$ and $-1.29$, while for the Pantheon$^{+}$+DESI DR2+CC+RSD dataset we obtain $s_0 = -2.50$, highlighting a stronger sensitivity to the dataset combination.

Overall, despite some quantitative differences between the BA and JBP parametrizations particularly at higher orders the qualitative behavior of all cosmographic parameters remains consistent across datasets. The inclusion of cosmic chronometer and RSD data induces only moderate shifts and does not alter the global evolutionary trends. These results demonstrate that both parametrizations provide a stable and coherent description of the late-time cosmic expansion up to higher order derivatives, while allowing for mild deviations from the standard $\Lambda$CDM model, especially in the detailed behavior of $j(z)$ and $s(z)$.

\begin{figure*}
    \centering
    \includegraphics[width=0.5\linewidth]{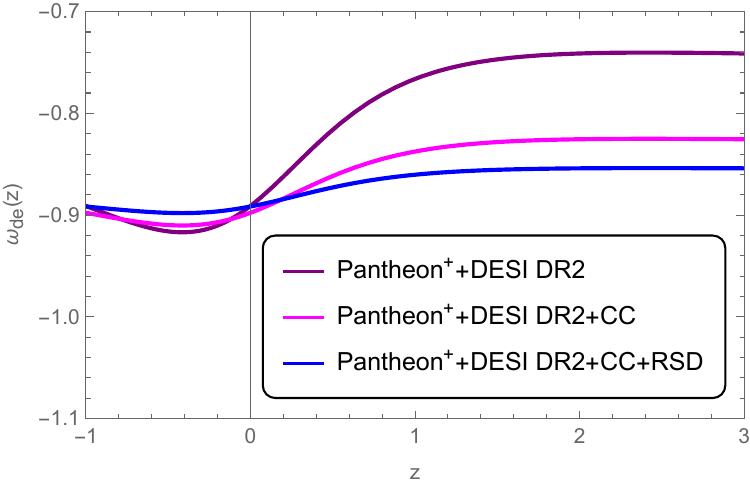}\includegraphics[width=0.5\linewidth]{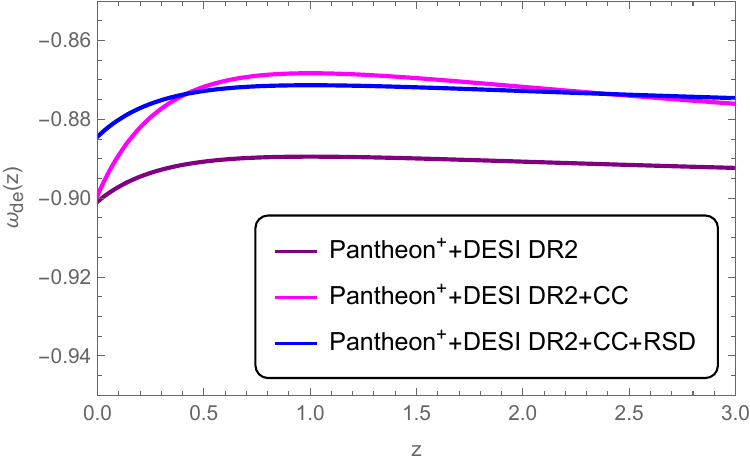}
    \caption{The evolution of the equation of state $\omega_{de}$(z) for the BA model (left panel) and the JBP model (right panel) as functions of redshift.}
    \label{fig:8}
\end{figure*}

Fig~\ref{fig:8} illustrates the redshift evolution of the dark energy equation of state parameter, $\omega_{de}(z)$, for both the BA (left panel) and JBP (right panel) parametrizations, based on the Pantheon$^{+}$ combined with DESI DR2, and Pantheon$^{+}$+DESI DR2+CC datasets. In both scenarios, the evolution of $\omega_{de}(z)$ is smooth over the considered redshift range and remains within the quintessence regime at the present epoch ($z=0$) i.e. $\omega_{de} > -1$. More precisely, within the BA framework, the present day values are $\omega_{de}(0) = -0.891$ for the Pantheon$^{+}$+DESI DR2 dataset and $\omega_{de}(0) = -0.897$ when cosmic chronometers are included, both clearly indicating a quintessence-like behavior. Similarly, for the JBP parametrization, one finds $\omega_{de}(0) = -0.900$ and $\omega_{de}(0) = -0.899$ for the same respective datasets, again consistent with a non-phantom dark energy component. For the Pantheon$^{+}$+DESI DR2+CC+RSD dataset, we obtain $\omega_{de}(0) = -0.891$ for the BA model and $\omega_{de}(0) = -0.884$ for the JBP parametrization, confirming that the present day equation of state remains in the quintessence regime when RSD data are included. Interestingly, in all cases, a transition across the phantom divide line ($\omega_{de} = -1$) is predicted to occur in the future, corresponding to negative redshift values. For the BA model, this crossing takes place at $z \simeq -0.414$ for both dataset combinations, suggesting a stable prediction with respect to the inclusion of CC data, and remains unchanged at $z \simeq -0.414$ when RSD data are included. In contrast, the JBP parametrization exhibits a more pronounced sensitivity, with the crossing occurring at $z \simeq -1.94$ for Pantheon$^{+}$+DESI DR2 and shifting further to $z \simeq -3.72$ when CC data are included, while for the full dataset including RSD we obtain $z \simeq -1.934$. These findings point toward a dynamically evolving dark energy sector that remains consistent with current observational constraints, while allowing for a possible future transition into the phantom regime. Such behavior highlights mild but non negligible deviations from the standard $\Lambda$CDM model at late times, depending on the chosen parametrization and dataset combination.

\subsection{Om diagnostic}

\begin{figure*}
    \centering
    \includegraphics[width=0.5\linewidth]{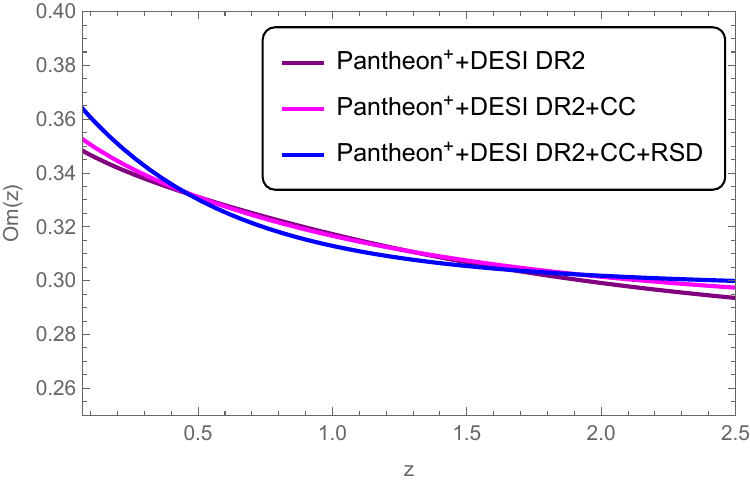}\includegraphics[width=0.5\linewidth]{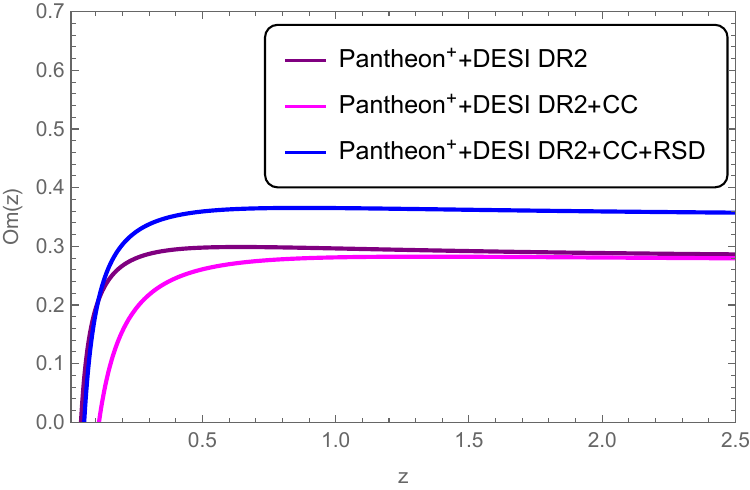}
    \caption{The evolution of the Om(z) diagnostic for the BA model (left panel) and the JBP model (right panel) as functions of redshift.}
    \label{fig:10}
\end{figure*}

The $\mathrm{Om}(z)$ diagnostic provides a simple yet powerful tool for distinguishing the standard $\Lambda$CDM model from scenarios involving dynamical dark energy. Its main advantage lies in its direct dependence on the Hubble parameter, which makes it relatively robust against observational uncertainties. In a spatially flat Universe, it is defined as
\begin{equation}
\mathrm{Om}(z) = \frac{E^2(z)-1}{(1+z)^3-1}.
\end{equation}
As shown in Fig.~\ref{fig:10}, the reconstructed $\mathrm{Om}(z)$ curves for the Pantheon$^{+}$+DESI DR2, Pantheon$^{+}$+DESI DR2+CC, and Pantheon$^{+}$+DESI DR2+CC+RSD datasets lie below the constant $\Lambda$CDM value, indicating a deviation from a strictly constant dark energy component.

For the BA parametrization, the $\mathrm{Om}(z)$ curves remain systematically below the $\Lambda$CDM value and exhibit a negative slope, indicating an effective phantom-like behavior. The combination of multiple datasets, namely Pantheon$^{+}$+DESI DR2+CC, leads to tighter constraints and a smoother evolution. For the Pantheon$^{+}$+DESI DR2+CC+RSD dataset, the BA parametrization preserves the same qualitative behavior, with the $\mathrm{Om}(z)$ curve remaining below the $\Lambda$CDM value and maintaining a negative slope, confirming the robustness of the phantom-like signature when RSD data are included.

In contrast, the JBP parametrization exhibits a different behavior, consistent with a quintessence-like regime. In particular, at low redshift, $\mathrm{Om}(z)$ shows a rapid increase before gradually approaching a nearly constant value at higher redshift. This trend reflects a stronger dynamical evolution of the dark energy component compared to the BA case, without indicating a phantom behavior. For the full dataset including RSD, the JBP model maintains this characteristic evolution, with only moderate quantitative changes, indicating that the inclusion of RSD data does not alter its overall dynamical pattern.

The noticeable separation between the curves at low redshift indicates an enhanced sensitivity to the inclusion of CC data, although both datasets converge toward similar values at earlier times, suggesting a reduced dependence on observational inputs in the high redshift regime. This behavior remains consistent when RSD data are added, with slightly tighter constraints but no significant modification of the global trends.

From a theoretical perspective, these results reinforce the interpretation that the underlying $f(Q)$ framework effectively mimics a dynamical dark energy sector whose behavior departs from a cosmological constant. In particular, the persistent negative slope in the BA case points toward an effective equation of state in the phantom domain without invoking fundamental phantom fields, highlighting the role of modified gravity in generating such behavior geometrically. Meanwhile, the JBP parametrization, with its quintessence-like behavior and pronounced redshift dependence, suggests a richer dynamical structure that may be associated with higher-order contributions in the gravitational sector or evolving effective degrees of freedom. Consequently, the $\mathrm{Om}(z)$ diagnostic not only provides a robust consistency check against $\Lambda$CDM, but also offers deeper insight into the underlying gravitational dynamics, supporting the viability of power law $f(Q)$ models as a compelling alternative for describing late-time cosmic acceleration.

\section{Conclusion}
\label{sec:6}

In this work, we have constrained the considered cosmological models using a combination of Pantheon$^{+}$ Type Ia supernovae, DESI DR2, cosmic chronometer (CC), and RSD datasets. This joint analysis allows us to probe the expansion history of the Universe over a wide redshift range and to obtain robust constraints on the model parameters.

For the $\Lambda$CDM model, we obtain stable and consistent results across the different dataset combinations. In particular, using the Pantheon$^{+}$+DESI DR2+CC dataset, we find $\Omega_m = 0.3075 \pm 0.0099$ and $h = 0.735 \pm 0.011$, together with $M = -19.253 \pm 0.031$ and $r_d = 137.1 \pm 2.1$. These values remain fully consistent with the standard cosmological scenario. When including RSD data, the constraints remain essentially unchanged, with $\Omega_m = 0.3069 \pm 0.0098$ and $h = 0.736 \pm 0.010$, while providing an additional constraint on the growth parameter $\sigma_8 = 0.791 \pm 0.024$.

For the BA parametrization, the Pantheon$^{+}$+DESI DR2+CC dataset yields $\Omega_m = 0.292 \pm 0.015$ and $h = 0.732 \pm 0.010$, with dark energy parameters $\omega_0 = -0.895 \pm 0.046$ and $\omega_1 = -0.012^{+0.097}_{-0.076}$. The additional parameters are constrained to $n = -1.44^{+0.14}_{-0.19}$ and $\gamma = 0.183^{+0.014}_{-0.021}$, while $r_d = 136.2 \pm 2.1$. When RSD data are included, the results remain broadly consistent, with $\Omega_m = 0.287^{+0.017}_{-0.015}$ and $h = 0.733 \pm 0.011$, $\omega_0 = -0.887 \pm 0.048$ and $\omega_1 = 0.023^{+0.069}_{-0.078}$, while $n = -1.110 \pm 0.063$ and $\gamma = 0.222^{+0.011}_{-0.014}$, and $\sigma_8 = 0.819 \pm 0.031$.

Similarly, for the JBP parametrization, we obtain for the Pantheon$^{+}$+DESI DR2+CC dataset $\Omega_m = 0.261^{+0.022}_{-0.019}$ and $h = 0.734 \pm 0.011$, with $\omega_0 = -0.889 \pm 0.052$ and $\omega_1 = 0.267 \pm 0.051$. The parameters $n = -1.05^{+0.17}_{-0.12}$ and $\gamma = 0.240^{+0.025}_{-0.020}$ are well constrained, with $r_d = 136.1 \pm 2.2$. With the inclusion of RSD data, the constraints shift to $\Omega_m = 0.314^{+0.029}_{-0.035}$ and $h = 0.710^{+0.035}_{-0.030}$, with $\omega_0 = -0.889 \pm 0.049$ and $\omega_1 = -0.065^{+0.091}_{-0.072}$, while $n = -1.02^{+0.11}_{-0.13}$ and $\gamma = 0.228^{+0.020}_{-0.027}$, and $\sigma_8 = 0.810 \pm 0.030$. In both parametrizations, the results indicate a dark energy behavior consistent with the quintessence regime at the present epoch, exhibiting a mild dynamical evolution.

The inclusion of Redshift Space Distortion (RSD) data provides an important test of structure formation and allows us to probe the effective gravitational coupling. In this context, the evolution of the ratio $G_{\mathrm{eff}}/G$ remains very close to unity over the redshift range $0 \leq z \leq 3$, typically satisfying $0.9993 \lesssim \frac{G_{\mathrm{eff}}}{G} \lesssim 1$, indicating only small deviations from General Relativity. This confirms that the growth of cosmic structures in the considered models closely follows the standard cosmological behavior.

From a statistical point of view, we find that both BA and JBP models slightly improve the fit to the observational data compared to $\Lambda$CDM, with $\Delta \chi^2 \simeq -4.81$ and $-4.50$ for Pantheon$^{+}$+DESI DR2+CC, and $\Delta \chi^2 \simeq -3.90$ and $-3.95$ when RSD data are included. However, once the number of free parameters is taken into account through the information criteria, the situation changes. The values $\Delta \mathrm{AIC}_c \simeq 1.23$ (BA) and $1.54$ (JBP) for Pantheon$^{+}$+DESI DR2+CC, and $\Delta \mathrm{AIC}_c \simeq 2.15$ (BA) and $2.10$ (JBP) for the full dataset, indicate that both models remain competitive with $\Lambda$CDM. In contrast, the larger $\Delta \mathrm{BIC}$ values, equal to $17.58$ and $17.89$ for Pantheon$^{+}$+DESI DR2+CC, and increasing to $18.52$ and $18.47$ when RSD data are included, reflect the stronger penalty associated with the increased model complexity. Overall, although the extended parametrizations provide a slightly better fit at the level of $\chi^2$, $\Lambda$CDM remains statistically preferred according to the BIC criterion. Nevertheless, the BA and JBP models offer viable alternatives, capable of capturing mild deviations from the standard scenario and describing a dynamical dark energy component.

Future investigations could further test these models by including additional observational probes, such as large-scale structure or weak lensing data, and by exploring possible constraints from local gravity tests. Such analyses would provide a deeper understanding of the nature of dark energy and the validity of extended cosmological models.

\bibliographystyle{compj}

\end{document}